# Synthesis of Cu mono-component metallic glass by the deposition on amorphous SiO$_2$ substrate: a molecular dynamics study

Yang Yu*

*Nanjing University of Information Science and Technology, School of Physics and Optoelectronic Engineering, China*

*Authors to whom correspondence should be addressed: Dr. Yang Yu, Electronic mail: yuyang@nuist.edu.cn

**Abstract:**

In this work, we simulated the physical vapor deposition (PVD) process of Cu atoms on the amorphous SiO$_2$ substrate. The resulting Cu thin layer exhibit amorphous structure. The Cu liquid quenching from 2000 K to 50 K was also simulated with different cooling rate to form the Cu metallic glass for comparison. The Cu glasses from the two different processes (PVD and quenching) revealed the same radial distribution function but different local structure from the Voronoi tessellation analysis. The PVD glass exhibit higher densities and lower potential energy compared with the melt-quenched counterpart, which corresponded to the properties of ultrastable glasses.

Keywords: mono-component metallic glass; ultrastable metallic glass; molecular dynamics simulation; physical vapor deposition.

## 1. Introduction:

Metallic glasses that have been first fabricated in 1960[1], exhibit a number of attractive properties (high hardness, high corrosion resistance, good ductility and toughness) [2-6]. Although the glassy state is a universal property of supercooled glasses[7], due to the low-glass forming ability, the vitrification of metallic liquids is notoriously difficult. Very high quenching rate ($10^5$ ~ $10^6$ K/s) which is about $10^5$ times of the typical laboratory cooling rate (1 K/s) is necessary to solidify a melt before the crystallization takes place for the binary, ternary and multicomponent alloys. As for the pure metals, ultrahigh cooling rate (at least $10^{12}$ K/s) is required. Zhong et al. fabricated tantalum and vanadium monatomic metallic glasses with the cooling rate of $10^{14}$ K/s [8, 9]. The cooling rate of individual atoms precipitating onto a cold substrate can reach to as high as $10^{10}$ ~ $10^{14}$ K/s. Stella et al. obtained tantalum metallic glass film by e-beam evaporators[10]. As the simplest glass formers, monatomic metallic glasses provide original aspects to study the properties and structure relationship of glasses. In this work, the Cu atoms deposited on SiO$_2$ substrate was modeled; thin film with amorphous structure was resulted. The constructed thin film exhibit higher density and lower potential energy compared with its supercooled counterpart, which is representative for the ultrastable glass. This result demonstrates that the ultrastable glass properties is ubiquitous for organic glasses[11, 12], multi-component metallic glasses[13-15], as well as monatomic metallic glasses - the most simple glass.

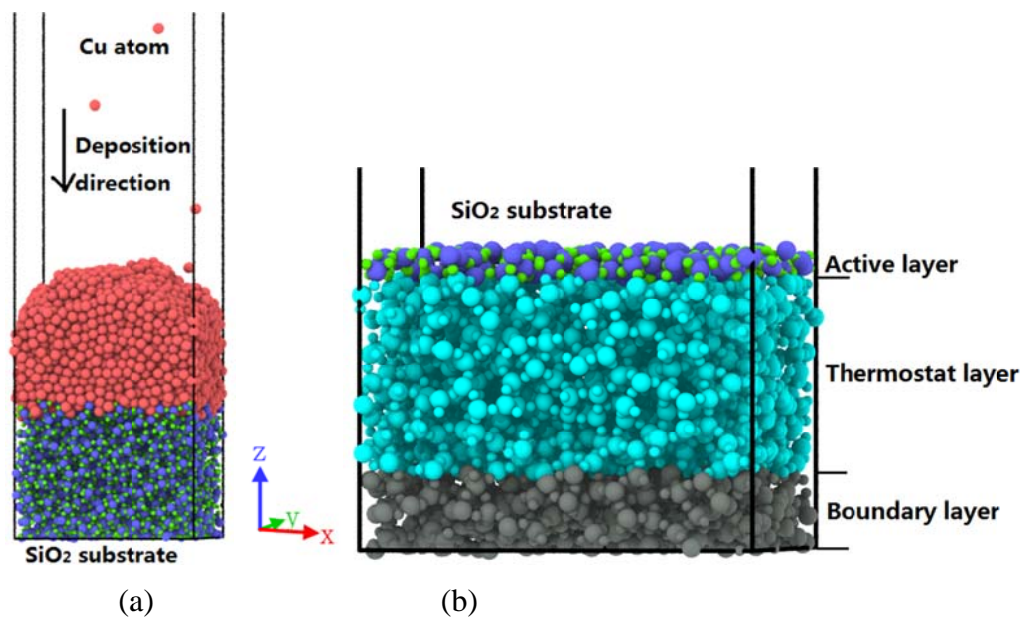

**Fig. 1.** (a) The snapshot of Cu thin film deposited on amorphous $SiO_2$ substrate with T = 300 K. (b) Layered configuration of the amorphous $SiO_2$ substrate.

2. **Simulation method:**

All molecular dynamics (MD) simulations were performed using the LAMMPS (large-scale atomic/molecular massively parallel simulator) software package [16, 17]. A time step of 1 fs was applied in all cases. The software OVITO [18, 19] was employed for the visualization of the simulation results as well as the analysis of radial distribution function (RDF) and Voronoi tessellation[20].

2.1 Preparation of amorphous $SiO_2$ surfaces

Amorphous $SiO_2$ substrate was prepared by cleaving of a bulk amorphous $SiO_2$, which was prepared from supercooled liquid $SiO_2$. A Tersoff-type empirical interatomic potential [21] was selected for the Si-O system in this work. First the α-quartz $SiO_2$ crystal lattice containing 24300 atoms (8100 Si atoms and 16200 O atoms) of the dimension 44.2443 Å × 42.5729 Å × 162.162 Å was created under periodic boundary conditions in all three directions. The system was melted at 5000 K for 5 ns, and then quenched from 5000 K to 300 K at a cooling rate of $18.8 \times 10^{12}$ K/s under NPT (constant number, pressure and temperature) ensemble. The amorphous bulk is of the resulting density 2.3 g/cm$^3$, in accordance with other simulation studies [22, 23] and is about 5% higher than the experimental data[24]. To obtain the substrate with surface lying parallel to the X-Y plane, the periodic boundary condition in the Z direction was removed. The bulk was sliced in the X-Y plane, leaving a slab with the dimension 44.2443 Å × 42.5729 Å × 30 Å, consisting of 4781 atoms.

The amorphous $SiO_2$ substrate as showed in Fig. 1b was divided into three layers: The bottom atoms of 5 Å thick constructed the boundary layer; this bottom layer was used to prevent atom loss. The middle layer of 22 Å thick was thermostat layer; the atoms in the middle layer were maintained at a constant temperature of 300 K by means of velocity rescaling thermostat [25-28]. The top layer of 3 Å thick was active layer. The atoms in the middle and top layers were then evolved in the microcanonical NVE (constant number N, volume V and energy E) ensemble.

2.2 Cu deposition simulation

The COMB (charged optimized many-body) potential [29] was applied to model the Cu-Si and Cu-O short-range interactions; the realistic EAM (embedded-atom-method) interaction potential parameterized by Mendelev et al. [30] was utilized to describe the atomic interaction of Cu atoms; and the tersoff type potential [21] was applied for the $SiO_2$ substrate. Cu atoms were introduced into the region above the $SiO_2$ substrate with randomly chosen X and Y coordinates, while the Z-coordinate of each inserted atom was 130 - 132 Å above the slab. The deposited atoms were initialized with the downward velocity randomly distributed from 480 m/s to 530 m/s. The deposition rate was one atom per 4 ps. The simulation was carried out within the microcanonical NVE ensemble for the deposited Cu atoms and the atoms of the $SiO_2$ substrate. The thermostat layer of the $SiO_2$ substrate maintained a constant temperature of 300 K by the velocity rescaling thermostat [25-28]. The snapshot of the deposition process was shown in Fig.1a. After the deposition procedure, the Cu deposition layer was separated from the substrates to eliminate the effect of the $SiO_2$ layer when analyzing the physical properties and this Cu film was denoted as PVD300 in this work. When calculating the density and potential energy, the data of the middle region of the film were chosen to remove the surface effect.

2.3 Supercooled Cu bulk glass

For comparison, the Cu liquid quenching from 2500 K to 50 K with different high cooling rate (from $10^{11}$ K/s to $10^{14}$ K/s) was also modeled. The same EAM potential [30] as for the deposition Cu film was applied for the interaction of the supercooled Cu atoms. The FCC crystal cubic with periodic boundary condition in three dimensions was initialized with 54.2235 Å length of each side. The simulation cell was comprised of 13500 atoms. The sample was heated from 300 K to 2500 K under the NPT ensemble. The temperature 2500 K was chosen higher than the melting temperature in a way to reach equilibrium disordered liquid state. After relaxing the structure in the liquid state for 50 ps with NVT (constant number, volume and temperature) ensemble, we have then cooled down the system from 2500 to 50 K with different cooling rates using NPT ensemble. When comparing with the PVD300 sample, the quenched samples were analyzed under 300 K.

3. **Result and discussion:**

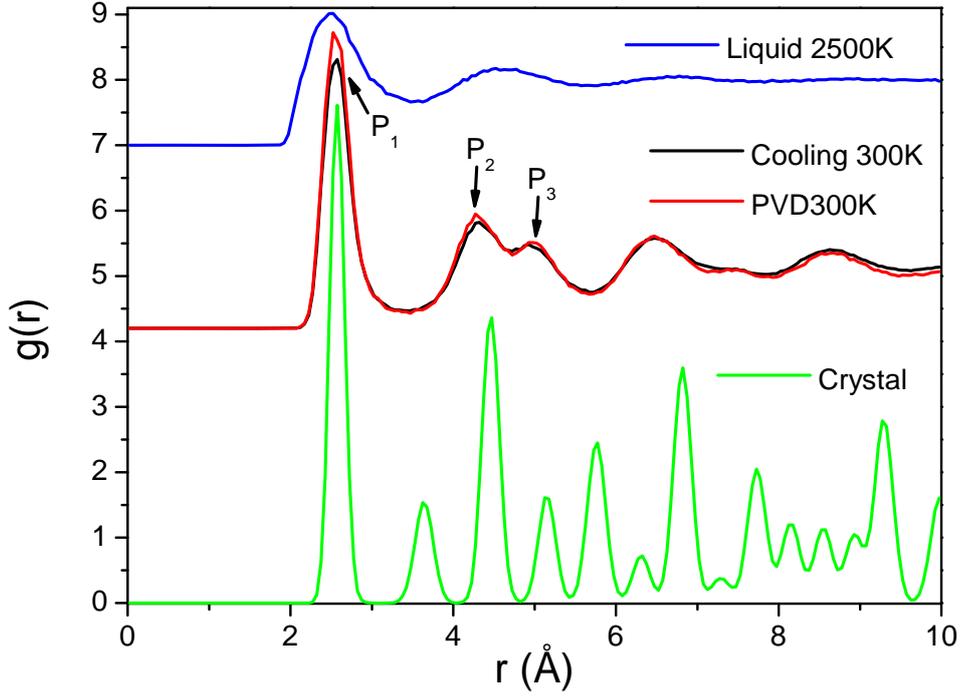

**Fig. 2.** The radial distribution functions (RDF) of liquid Cu at 2500 K, supercooled Cu at 300 K, deposited Cu glass (PVD300K) and crystal Cu at 300 K. The $P_i$ (i=1,2,3) denotes the RDF peak positions of Cu metallic glass samples.

The radial distribution function (RDF) results of Cu crystal at 300 K, PVD300, supercooled Cu glass with cooling rate $10^{13}$ K/s at 300 K and liquid Cu at 2500 K are shown in Fig. 2. The lower green line which correspond to the Cu crystal reveals the long range order characteristic of the crystal structure. The red line denoted as PVD300K represents the Cu thin film prepared from deposition. The RDF of PVD300K is almost identical to the Cooling 300 K (black line) which represents the supercooled glass. It shows amorphous structure which has no long range order in the atomic configuration. The peak positions of the RDF of PVD300K are: $P_1$=2.525 Å, $P_2$=4.275 Å, $P_3$=4.925 Å. The position ratio $P_2/P_1$ is 1.69, and $P_3/P_1$=1.95. These values are close to $\sqrt{3}$ (=1.73) and $\sqrt{4}$ (=2), which is a general character for the metallic glasses [31-33] and in agreement with the simulation result of the reported work [34]. There is a split between the second and third peaks of the RDF of Cu metallic glass, which is characteristic of the amorphous solids [35]. This split vanishes in the RDF of liquid Cu as shown in the blue line (Liquid 2500 K) and replaced by one pronounced peak with a gentle slope. The splitting of the second peak of the melt into two sub-peaks of the corresponding amorphous solid in the RDF is attributed to the appearance of the icosahedra cluster [36]. From the topographic view, the splitting is caused by the addition of the local translational symmetry to the spherical-periodic order [32].

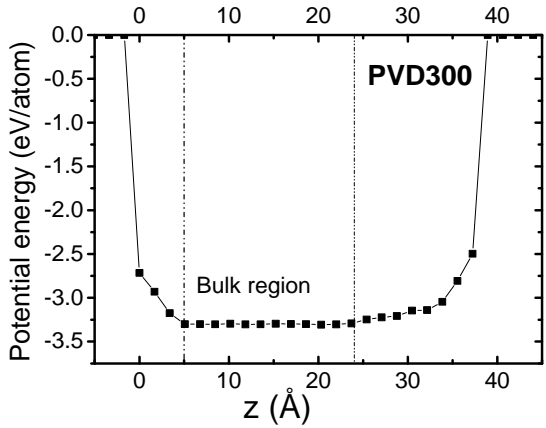

**Fig. 3.** Potential energy profile of the PVD300 thin film along the Z-direction. The vertical dash lines are the boundaries of the bulk region.

The local potential energy of the PVD300 film along the growth direction Z was shown in Fig.3. Because of the surface energy effect as well as the unevenness of the atom distribution on the surface, the average potential energy exhibits lower value near the two surfaces. There is a plateau region in the middle about 5 Å to 24 Å above the lower surface, with the thickness of 19 Å. This region is denoted as bulk region and the potential energy in this region was chosen as the potential energy of the PVD300 film. The average potential energy is -3.302 eV. The density value of PVD300 was also carried out within this central region and the value is 8.52 g/cm$^3$, which is higher than the supercooled Cu bulk with the value 8.50 g/cm$^3$. The density of Cu crystal structure in this simulation work was 8.70 g/cm$^3$, this structure is modeled under the pressure of 0 Pa and temperature of 300 K. The density of Cu glass structure is lower than the crystal counterpart.

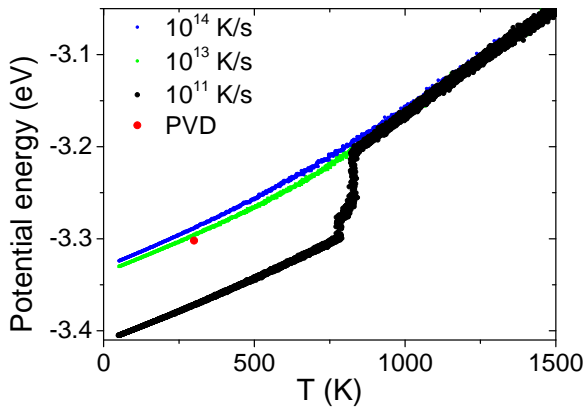

**Fig. 4.** Potential energy as a function of temperature for Cu melt during the process of quenching at three cooling rates ($10^{11}$ K/s, black dot; $10^{13}$ K/s, green dot; $10^{14}$ K/s, blue dot). The red dot represents the potential energy of PVD300.

Fig. 4 shows the temperature dependence of the potential energy per atom during cooling process with various cooling rates $10^{11}$ K/s, $10^{13}$ K/s and $10^{14}$ K/s. The potential energy of PVD300 at 300 K is also shown in the figure as the red dot. During the quenching process of cooling rate $10^{11}$ K/s, the potential energy drops suddenly in a discontinuous way, corresponding to the first-order phase transition as a result of the crystallization. There is no abrupt drop for the samples of the cooling rate $10^{13}$ K/s and $10^{14}$ K/s, indicating the glass-formation in these systems. The potential energy of crystalline structure is obviously lower than the glass structure. For the supercooled samples, the slower the cooling rate, the lower the potential energy, which corresponds to a more stable structure. The potential energy of PVD300 falls below the supercooled glass and high above the crystal structure.

To estimate the cooling rate corresponding to the potential energy of PVD300, we sampling the potential energy at 300 K of the Cu glasses with various cooling rates ($5\times10^{12}$ K/s, $8\times10^{12}$ K/s, $10^{13}$ K/s, $2\times10^{13}$ K/s, $5\times10^{13}$ K/s and $10^{14}$ K/s) and draw the picture as Fig. 5. The quenching samples were relaxed at 300 K for 1 ns and average the potential energy for the next 1 ns. The dashed line is the fitting result of the data with a logarithmic function. Extrapolating the fitting line to the line corresponding to the potential energy (-3.302 eV) of the PVD sample at 300 K, we estimate that the cooling rate corresponding to PVD300 is about $6\times10^9$ K/s. This cooling rate is lower than the crystallization rate $10^{11}$ K/s of Cu liquid and at least 3 orders of magnitude slower than the supercooled glass available in this study. From the result in this study, deposited Cu glass exhibit higher density and lower potential energy which corresponding to a much lower cooling rate than the ordinary supercooled glass, these are typical characters of ultrastable glasses.

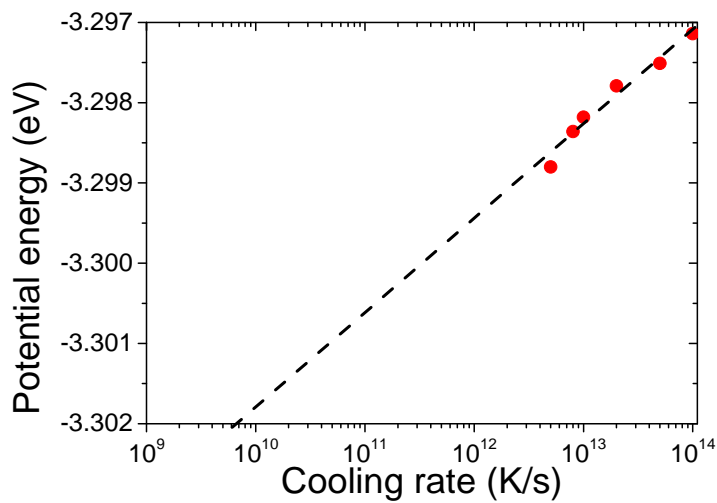

**Fig. 5.** Potential energy per atom at 300 K for melt-quenched samples as a function of different cooling rates: $5\times10^{12}$ K/s, $8\times10^{12}$ K/s, $10^{13}$ K/s, $2\times10^{13}$ K/s, $5\times10^{13}$ K/s and $10^{14}$ K/s. The dashed line is a logarithmic function to fit the data. We extrapolated the line to the potential energy (-3.302 eV) of the PVD sample at 300 K. The estimated cooling rate corresponding to the potential energy is $6\times10^{10}$ K/s.

Table 1. Fraction and LFFS of the most domimant Voronoi Polyhedrons in the supercooled (CR13) and deposited (PVD300) Cu metallic glass.

| Voronoi polyhedra | | CR13 | PVD300 | LFFS |
|---|---|---|---|---|
| Icosahedra-like | <0,0,12,0> | 4.9 | 5.9 | 1 |
| | <0,0,12,2> | 1.8 | 2.6 | 0.86 |
| | <0,1,10,2> | 9.3 | 10.5 | 0.77 |
| | <0,1,10,3> | 4.8 | 5.8 | 0.71 |
| | <0,1,10,4> | 2.8 | 3.1 | 0.67 |

|  | Voronoi index | CR13 | PVD300 |  |
|---|---|---|---|---|
|  | <0,2,8,2> | 1.8 | 2.0 | 0.67 |
|  | <0,2,8,3> | 2.4 | 1.8 | 0.62 |
|  | <0,2,8,4> | 10.5 | 10.1 | 0.57 |
|  | <0,2,8,5> | 4.1 | 4.5 | 0.53 |
|  | sum | 42.4 | 46.3 |  |
| Mixed-like | <0,3,6,4> | 7.8 | 7.4 | 0.46 |
|  | <0,3,6,5> | 6.2 | 5.3 | 0.43 |
|  | <0,3,6,6> | 3.0 | 2.7 | 0.4 |
|  | sum | 17 | 15.4 |  |
| Crystal-like | <0,4,4,6> | 4.1 | 3.3 | 0.29 |
|  | <0,4,4,7> | 1.6 | 1.4 | 0.27 |
|  | sum | 5.7 | 4.7 |  |

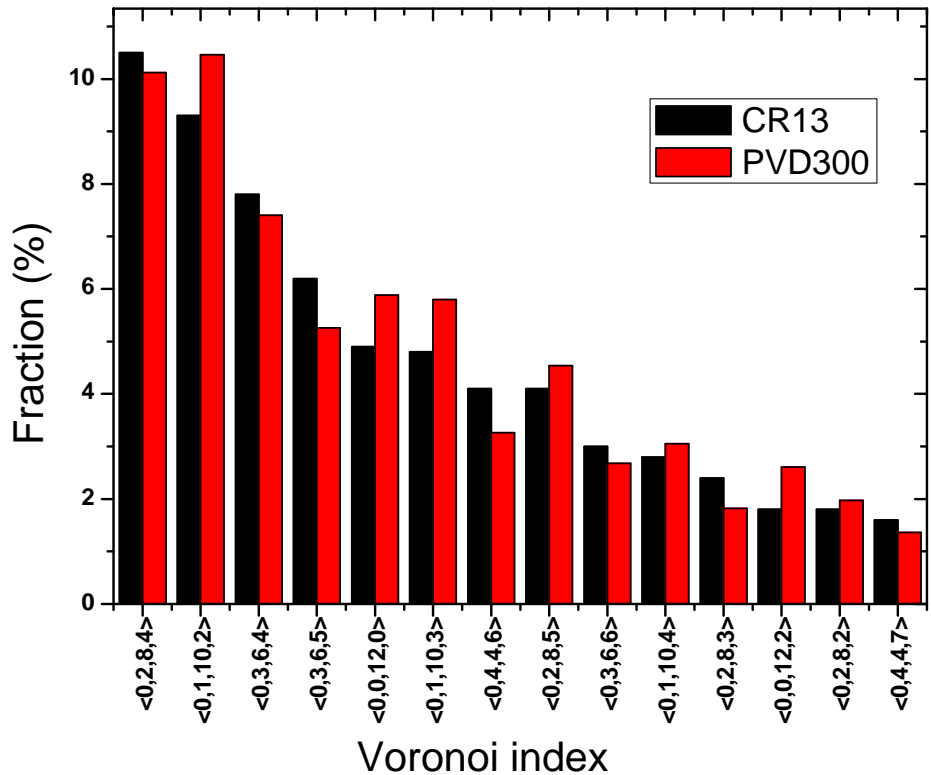

**Fig. 6.** Fractions of the most populous Voronoi polyhedrons in Cu metallic glass at 300 K (red: PVD300; black: CR13, supercooled Cu glass with cooling rate $10^{13}$ K/s).

The local atomic structure of the samples was characterized by means of Voronoi tessellation analysis in this work. The CR13 represents supercooled Cu glass with cooling rate $10^{13}$ K/s. Four indices $<n_3,n_4,n_5,n_6>$ designate different Voronoi Polyhedrons (VP) [20]. Here $n_i$ denotes the number of i-edged (i=3,4,5,6) faces of the polyhedron. The distribution of the most dominant Voronoi polyhedrons in the Cu glasses was shown in Fig.6. Unlike the multicomponent systems, whose dominant topological cluster is fully-icosahedral <0 0 12 0> [37, 38], in the pure metallic glass, fully-icosahedral is classified the fourth and fifth abundant structure in PVD300 and CR13 respectively and the dominant Voronoi polyhedrons are distorted icosahedra clusters <0,2,8,4> and <0,1,10,2>. This result is in accordance with the other work of Cu, Al and Pt monatomic metallic glass [39-41]. According to the work of Hwang et al. [42], Voronoi polyhedrons can be classified into three categories: icosahedral-like clusters (consisting <0,0,12,x>,<0,1,10,x> and <0,2,8,x>), mixed-like clusters (consisting <0,3,6,x> and <0,3,7,x>) and crystal-like clusters (consisting <0,4,4,x>), here x denotes a whole number from 0 to 7. The Voronoi polyhedral abundance in PVD300 and CR13 shows systematic correlation with the Voronoi polyhedral categories: the mixed-like (<0,3,6,x>, x=4,5,6) and crystal-like (<0,4,4,x>, x=6,7) clusters show lower fraction in PVD300 than in CR13, while the majority of the icosahedral-like (fully-ico <0,0,12,0>; distorted icosahedra <0,0,12,2>, and <0,1,10,x>, x=2,3,4; <0,2,8,x>, x=2,5) clusters show higher fraction in PVD300 than in CR13.

The degree of local fivefold symmetry (LFFS) of Voronoi polyhedrons is defined as $d_5 = n_5/\sum_i n_i$. Here, $n_5$ denotes the number of pentagons of Voronoi polyhedron, $n_i$ denotes the number of i-edged (i=3,4,5,6) faces of the polyhedron. The variance of the fraction shows divisions for Voronoi polyhedrons in PVD300 and CR13. From table 1, Voronoi polyhedrons with high degree of LFFS (<0,0,12,0>, <0,0,12,2>, <0,1,10,2> and <0,1,10,3>) shows higher abundance in PVD300 than CR13, while the fraction of low degree of LFFS (mixed-like and crystal-like Voronoi polyhedrons) shows a lower value in the PVD300 than CR13. For the metallic glasses, higher degree of LFFS corresponds to lower potential energy and lower free volume [43]. The Voronoi tessellation result is consistent with the former potential and density result and confirms the PVD300 sample is in ultrastable state.

## 4. Conclusion:

We carried out MD simulation for the physical vapor deposition of Cu metallic glass. The PVD glass shows higher density and lower potential energy than the melt-quenched counterpart. In order to create the glass by quenched technique with the same potential energy as the PVD sample, the cooling rate should be at least 3 orders of magnitude slower than the supercooled glass available in this study, which is lower than the cooling rate of crystallization of Cu liquid. The increased stability of PVD sample is correlated to the local structure feature from Voronoi tessellation analysis. The local structure of PVD glass exhibits the higher degree of local fivefold symmetry compared with the quenched counterparts.

*Acknowledgements*

*Dr. Yang Yu acknowledges the financial support from the National Natural Science Foundation of China (Grant No. 11704194).*